\documentclass[aps,pra,superscriptaddress,notitlepage,floatfix,notitlepage, twocolumn]{revtex4-2}
\usepackage[utf8]{inputenc}
\usepackage{graphicx}
\usepackage{float}
\usepackage{tikz}
\usepackage{amsmath}
\usepackage{amssymb}
\usepackage[normalem]{ulem}

\usepackage{multirow}
\usepackage[colorinlistoftodos]{todonotes}
\usepackage[colorlinks=true, allcolors=black]{hyperref}

\newcommand{\ie}{i.\,e.}%

\makeatletter
\def\subsubsection{\@startsection{subsubsection}{3}{10pt}{-1.25ex plus -1ex minus -.1ex}{0ex plus 0ex}{\normalsize\bf}}
\def\paragraph{\@startsection{paragraph}{4}{10pt}{-1.25ex plus -1ex minus -.1ex}{0ex plus 0ex}{\normalsize\textit}}
\renewcommand\@biblabel[1]{#1}
\renewcommand\@makefntext[1]%
{\noindent\makebox[0pt][r]{\@thefnmark\,}#1}
\DeclareRobustCommand\onlinecite{\@onlinecite}
\def\@onlinecite#1{\begingroup\let\@cite\NAT@citenum\citealp{#1}\endgroup}
\def\tagform@#1{\maketag@@@{\ignorespaces#1\unskip\@@italiccorr}}
\let\orgtheequation\theequation
\def\theequation{(\orgtheequation)}
\makeatother

\begin{document}

\author{Juan M. García-Garrido}
\affiliation{Departamento de F\'{\i}sica At\'omica, Molecular y Nuclear,
 Universidad de Granada, 18071 Granada, Spain}
\author{Valery Milner}
\affiliation{Department of Physics and Astronomy, The University of British Columbia, Vancouver, Canada}
\author{Christiane P. Koch}
\affiliation{Freie Universität Berlin, Dahlem Center for Complex Quantum Systems and Fachbereich Physik, Germany}
\author{Rosario González-Férez}
\affiliation{Departamento de F\'{\i}sica At\'omica, Molecular y Nuclear,
 Universidad de Granada, 18071 Granada, Spain}
 \affiliation{Instituto Carlos I de F\'{\i}sica Te\'orica y Computacional,
 Universidad de Granada, 18071 Granada, Spain} 

\begin{abstract}

We investigate theoretically the ability of an optical centrifuge - a laser pulse whose linear polarization is rotating at an accelerated rate, to control molecular rotation in the regime when the rigid-rotor approximation breaks down due to  coupling between the vibrational and rotational degrees of freedom. Our analysis demonstrates that the centrifuge field enables controlled excitation of high rotational states while maintaining relatively low spread along the vibrational coordinate. We contrast this to the rotational excitation by a linearly polarized Gaussian pulse of equal 
spectral width and pulse energy which, although comparable to the centrifuge-induced rotation, is unavoidably accompanied by a substantial broadening of the vibrational wavepacket.
\end{abstract}

\title{Rotational excitation of molecules in the regime of strong ro-vibrational coupling: Comparison between an optical centrifuge and a transform-limited pulse}

\section*{}
\vspace{-1cm}

\maketitle
\section{Introduction}
\label{sec:intro}

Laser pulses have long been used for controlling the rotation of molecules
~\cite{Stapelfeldt2003,Ohshima2010,Fleischer2012,KochRMP19}. Among multiple approaches to rotational control, the method of an optical centrifuge proved to be most successful in spinning molecules to extremely high rotational states, known as molecular super-rotors \cite{Karczmarek1999,Villeneuve2000}. The centrifuge is a linearly polarized laser pulse, whose polarization vector rotates with a constant angular acceleration. An interaction of the laser-induced dipole moment with the applied laser field results in a torque, which forces the molecule to follow the accelerated rotation of the field polarization. 

On the experimental side, optical centrifuges have been used in numerous studies of molecular structure and molecular dynamics (for recent reviews, see Refs.~\citenum{MacPhail2020, Mullin2025}). 
In these investigations, it has been assumed that the centrifuge drives the molecule up the rotational ladder of states, without explicitly driving Raman transitions up the vibrational manifold.
This approximation is well justified in the case of relatively light molecular species with 
strong molecular bonds and correspondingly high energies of Raman-active vibrational modes (i.e. $>1000$~cm$^{-1}$), which fall outside the available energy bandwidth of amplified femtosecond laser pulses ($\sim500$~cm$^{-1}$). 

In the case of heavier molecules, or those with softer molecular bonds, ro-vibrational coupling may significantly change the dynamics of a super-rotor. Previous theoretical studies have explored this scenario in the context of the centrifuge-induced dissociation~\cite{Li2000, Spanner2001, salas2023}, as well as the dissociation induced by a short laser pulse~\cite{LemehskoPRL2009}. Interestingly, the reverse process of creating a molecular bond can also be facilitated by the ro-vibrational coupling~\cite{Gonzalezferez2012}. On the experimental side, evidence of the ro-vibrational energy exchange in the field-free dynamics of molecules in extreme rotational states has been recently reported~\cite{Chen2023}. 

Dissociation is an extreme example of coupled ro-vibrational dynamics. In view of coherent control of ro-vibrational dynamics, it is also interesting to understand the spread of population across 
the vibrational manifold below the dissociation limit. One of the open questions is whether the optical centrifuge can provide an experimental tool to make the molecule climb the vibrational ladder in a controlled way (much like it does with the rotational ladder climbing). Alternatively, one may ask whether high rotational excitation of molecules with low vibrational energies can be executed with the optical centrifuge in such a way as to keep their vibrational state intact, thus protecting them from the potential dissociation.
These heavier molecules  may be susceptible to the vibrational excitation by the centrifuge field, leading to ro-vibrational coupling, which may significantly change the dynamics of a super-rotor.

Our aim here is to analyze theoretically the degree of the vibrational excitation, comparing the effect of the centrifuge with that of a simple Gaussian pulse.
To this end, we have carried out a full quantum 
mechanical analysis of the ro-vibrational 
dynamics of a diatomic molecule in a non-resonant laser field. We compare a pulse whose time envelope mimics one of the polarization axis of an optical centrifuge~\cite{Korobenko2014} with a Gaussian pulse with the same spectral bandwidth and carrying the same energy as the
centrifuge pulse. We have chosen Rb$_2$ as an example of a heavy diatomic molecule that can be prepared and studied in a trapped ultracold gas~\cite{LangPRL2008,DeissPRL2014,WolfPRL2019,HeScience2020}.

This article is organized as follows. 
In~\autoref{sec:system} we describe the system Hamiltonian including 
the interaction with the optical centrifuge and another non-resonant 
laser pulse. Section~\ref{sec:vibrational} is devoted to explore
the field-dressed dynamics of the rovibrational ground state
and of several excited states as well as of a thermal
sample. Conclusions and outlook are presented in~\autoref{sec:conclu}.


\section{System description and numerical method}
\label{sec:system}
We consider a diatomic molecule in the $a^3\Sigma^+$ electronic state exposed to a non-resonant laser pulse linearly polarized along the $Z$-axis of the laboratory-fixed frame (LFF).
Within the Born-Oppenheimer approximation, the nuclear Hamiltonian is given by 
\begin{equation}
{H}(t)={T}_R+\frac{\bold{N}^2}{2\mu R^2}+V(R)+H_{I}(t),
\label{eq:Htotal}
\end{equation}
where the first and second terms stand for the vibrational and rotational kinetic energies, 
respectively, $\bold{N}$ is the rotational angular momentum, $\mu$ the reduced mass of the 
molecule, $R$ the internuclear distance, and $V(R)$ the electronic potential energy 
curve. For a non-resonant laser field, the polarizability interaction reads
\begin{equation}
H_{I}(t)=-\frac{I(t)}{2c\varepsilon_0}\left(\Delta\alpha(R)\cos^2\theta+\alpha_\perp(R)\right),
\end{equation}
where $I(t)$ is the intensity of the laser field, $c$ the speed of light in vacuum, $\varepsilon_0$ the 
electric permitivity of vacuum, and $\theta$ is the Euler angle formed between the internuclear and laser 
polarization axes.
The polarizability anisotropy is $\Delta\alpha(R)=\alpha_\parallel(R)-\alpha_\perp(R)$, with
$\alpha_\parallel(R)$ and $\alpha_\perp(R)$ being the parallel and perpendicular components of the 
polarizability in the molecular-fixed frame (MFF).

The time-dependent Schrödinger equation associated with the Hamiltonian~\eqref{eq:Htotal} is solved by a grid representation of the wave function, i.e., the Fourier 
and discrete variable representations for the radial and angular coordinates, respectively~\cite{RonnieReview88,beck20001,LightCarrington}, and the time evolution is calculated using the Chebyshev propagator~\cite{RonnieReview94}. 
The field-dressed dynamics is analyzed in terms of the field-free rovibrational eigenstates  of the Hamiltonian~\eqref{eq:Htotal} (obtained by setting $t=0$), 
$\Phi_{\nu,N, M_N}(\bold{R})=\phi_{\nu,N}(R)Y_{N,M_N}(\theta,\varphi)$,
where $ \phi_{\nu,N}(R)$ and $Y_{N,M_N}(\theta,\varphi)$ are the vibrational and rotational parts, respectively,
with $Y_{N,M_N}(\theta,\varphi)$ being the spherical harmonics, and $(\nu,N,M)$ the vibrational, 
rotational, and  magnetic quantum numbers. 
For the sake of simplicity, we refer to the eigenstate $\Phi_{\nu,N, M_N}(\bold{R})$
by its quantum numbers $(\nu,N,M)$. 
For a given initial state $\Psi(\bold{R},t=0)=\Phi_{\nu_0,N_0,M_{N_0}}(\bold{R})$, 
the time-dependent wave function can be expressed as 
\begin{equation}
\Psi(\bold{R},t)=\sum_{\nu,N} C_{\nu_0, N_0,M_{N_0}} (\nu, N,t)\Phi_{\nu,N,M_N}(\bold{R},t),
\label{eq:expan}
\end{equation}
where the sum runs over all vibrational bands and rotational excitations 
up to a maximal value, which is chosen to ensure converged results. 
In this expansion~\eqref{eq:expan}, we have used that $M_{N_0}$ is a good quantum number due to 
the azimuthal symmetry. The coefficients $C_{\nu_0, N_0,M_{N_0}} (\nu, N,t)$ depend on time during the 
pulse duration. Afterwards, the Hamiltonian~\eqref{eq:Htotal} becomes time-independent, and these coefficients
acquire constant absolute values but varying phases. 
 
We consider two different laser fields, both linearly polarized along the LFF $Z$-axis.  
The first one is a ``two-dimensional'' centrifuge pulse (CP), which mimics the experimental pulses used in~\cite{Milner2016}. While the two-dimensional model greatly simplifies our analysis, we expect its result to apply to conventional three-dimensional centrifuge fields~\cite{Korobenko2014}, because our focus is on the effects which do not depend on the directionality of molecular rotation.
The intensity of this pulse is given by
\begin{equation}
\label{eq:centrifuge}
I_{C}(t)=
\begin{cases}
I_{C}^{0}\sin^2\left(\frac{\pi t}{2t_{0}}\right) g(t;\beta), &0\leq t\leq t_{0},\\
I_{C}^{0}g(t;\beta), &t_{0}<t\leq t_{C}-t_0,\\
I_{C}^{0}\sin^2\left(\frac{\pi(t_f-t)}{t_{0}}\right)g(t;\beta), &t_C-t_{0}<t\leq t_C, 
\end{cases}
\end{equation}
where $I_C^{0}$ is the peak intensity, $t_{0}$ the turning on 
and off time, and $t_C$ the duration of this pulse. 
The function $g(t;\beta)=\sin^2\left(\beta t^2\right)$ simulates 
the oscillatory behavior of an experimental optical centrifuge pulse, with the parameter $\beta$ 
being the analogue of the acceleration of the polarization axis rotation.
Here, the CP parameters are fixed to $\beta=0.3~$ps$^{-2}$, $t_0=3$~ps and $t_C=15$~ps.

The second pulse is a non-modulated Gaussian pulse (GP)
with the same spectral bandwidth as the CP, and intensity 
\begin{equation}
I_G(t)=I_G^0\exp\left(-\frac{\left(t-t_g\right)^2}{\sigma^2}\right),
\label{eq:gaussian}
\end{equation}
where $I_G^0$ is the peak intensity and
$2\sigma\sqrt{\log2}$ is the full width at half maximum (FWHM). 
This GP possesses the same spectral bandwidth as the considered CP if $\sigma$ is taken to be $\sigma=0.142~$ps$ $. 
In addition, we impose that both pulses carry the same energy, which happens when their peak intensities are related to one another as $I_G^0=24.05I_C^0$. The GP is centered at $t_g=0.671$~ps, 
twice the value of the FWHM of the GP, so 
that the intensity at $t=0$~ps is $\sim I_G^0\cdot10^{-10}$~W/cm$^2$. The duration of this 
pulse is $t_G=2t_g$~ps, \ie, four times the GP  FWHM. 
By fixing this value of $t_G$, we have ensured that the effects due to very small,
but non-zero, laser fields at the beginning and end of the
time-evolution are negligible.


\section{Field-dressed rovibrational dynamics}
\label{sec:vibrational}

We consider a Rb$_2$ molecule for which the lowest triplet electronic state $a^3\Sigma_u^+$ accommodates approximately
$41$ vibrational states with no rotational excitation, \ie, $N=0$. 
Note that due to the
molecular symmetries, the $a^3\Sigma_u^+$ potential only accommodates rotational states of even parity.
The number of bound rotational excitations decreases as the 
vibrational 
quantum number $\nu$ increases. For instance, the maximal values 
of the rotational quantum numbers are $N=152$, $5$ and $0$
for the vibrational bands $\nu=0,\, 39$ and $40$, respectively. 
For the lowest vibrational band of  Rb$_2$  in the  $a^3\Sigma_u^+$ electronic state, the rotational constant  is $0.0104$ cm$^{-1}$, whereas the vibrational splitting is $12.8$~cm$^{-1}$.

The interaction with the laser field, for moderate to strong intensities, is expected to lead to a significant hybridization of the rotational motion, followed by an impulsive alignment because their durations are 
significantly shorter than the Rb$_2$ rotational period of $\tau_B=1.6$~ns.
However, the field-induced dressed dynamics for the CP and GP should be different 
due to the very distinct time scales and ways in which the pulses transfer
the same energy to the molecule.

\subsection{Dynamics of the rovibrational ground state}
\label{sec:ground}

\begin{figure}
\includegraphics[width=0.95\columnwidth]{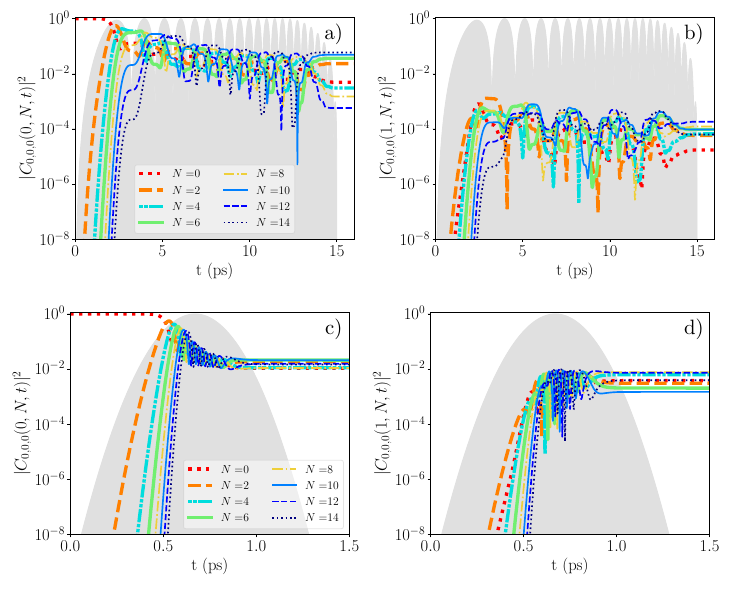}
\caption{For the initial state $(0,0,0)$, time evolution 
of the weights of the field-free states  $|C_{0,0,0}(\nu,N,t)|^2$ 
with rotational quantum number $N\leq 14$ and vibrational quantum numbers 
a) and c) $\nu=0$; and b) and d) $\nu=1$. 
The peak intensities are $I_G^0=10^{12}~$W/cm$^2$ and $I_C^0=4.158\cdot10^{10}~$W/cm$^2$. 
The shaded areas represent the time profiles of the pulses.}
\label{fig:evo}
\end{figure}
We assume that Rb$_2$ is initially in its rovibrational 
ground state $(0,0,0)$ and analyze the dressed dynamics 
by the time-evolution of
the projections of the wavepacket into the field-free basis,
\ie, $|C_{0,0,0}(\nu,N,t)|^2$,  presented in~\autoref{fig:evo}. 
We only plot the coefficients of the field-free states with 
$\nu=0,1$ and $N\le 14$. For the G and C pulses,
the peak intensities are fixed to 
$I_G^0=10^{12}~$W/cm$^2$ and $I_C^0=4.158\cdot10^{10}~$W/cm$^2$,
respectively.
Due to the selection rules $\Delta N=\pm2$ of the 
field interaction,  $(0,2,0)$ is 
the first state contributing to the field-dressed wavepacket,
 and immediately afterwards, higher rotational excitations are also populated. 
The weights $|C_{0,0,0}(0,N,t)|^2$ initially increase up to a maximum, 
keeping an oscillatory behavior till the end of the pulse when they reach 
a final constant value.
For the CP, the oscillatory behavior of $|C_{0,0,0}(0,N,t)|^2$ 
is a bit more irregular due to the consecutive maxima of $I_{C}(t)$.
The low values of the weights plotted in~\autoref{fig:evo} 
indicate that many 
rotational excitations contribute to the wave function.

An interesting phenomenon is that both pulses provoke that states within 
neighboring vibrational bands get populated. 
The comparison between panels (b) and (d) of~\autoref{fig:evo} shows that
their contribution is more important for the GP-dressed wavepacket, but even so,
they cannot be neglected for a proper description of the CP-induced
dynamics.
These non-zero weights of states with $\nu>0$
proves a strong ro-vibrational coupling between the vibrational and
rotational degrees of freedom, 
induced by the interaction with the non-resonant laser field, and the breakdown of the 
rigid-rotor approximation.
\begin{figure}
\includegraphics[width=0.95\columnwidth]{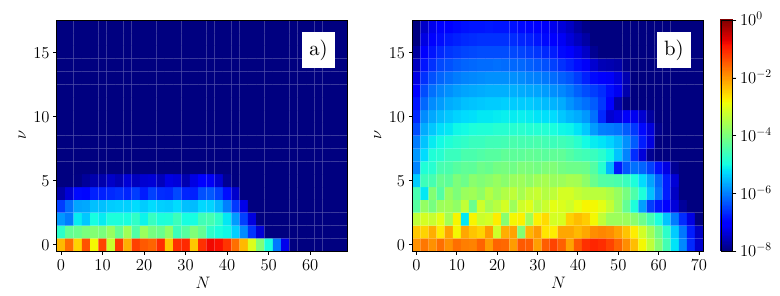}
\caption{For the initial state $(0,0,0)$, final weights of the field-free rovibrational states $|C_{\nu,N}^{0}(t_f)|^2$
 after the a) centrifuge and b) Gaussian pulses with  peak intensities $I_G^0=10^{12}~$W/cm$^2$ and $I_C^0=4.158\cdot10^{10}~$W/cm$^2$, respectively.}
\label{fig:2D_GS}
\end{figure}
The contribution of states from higher vibrational bands is also important at the end of the pulses, as shown by 
the weights of the field-free eigenstates plotted 
in~\autoref{fig:2D_GS}~(a)
and~\autoref{fig:2D_GS}~(b) for the G and C pulses, respectively.
These final weights illustrate the differences between the 
dynamics induced by these two fields. 
The distribution of population is wider in the vibrational and rotational quantum numbers for the GP-induced dynamics.
For instance, the neighboring vibrational bands
$\nu=1$ and $\nu=2$ have larger
cumulative weights, defined in~\autoref{eq:vib_dist}, 
${\cal V}_{0,0,0}(1,t=t_f)=0.160$ and
${\cal V}_{0,0,0}(2,t=t_f)=0.033$ due to the GP. 
In contrast, the vibrational distribution induced by the CP pulse is very narrow, with rather low 
cumulative weights, 
${\cal V}_{0,0,0}(1,t=t_f)=0.0018$ and ${\cal V}_{0,0,0}(2,t=t_f)=0.0003$,
which is due to its weaker peak intensity, and its gradual transfer of energy to the molecule.
Indeed, the centrifuge time envelope
enables a better control over the vibrational excitations that are not significantly increased
after the first intensity maximum.

\begin{figure}
\includegraphics[width=0.95\columnwidth]{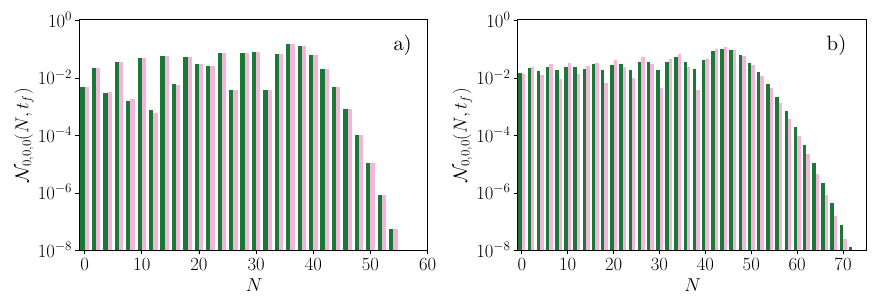}
\caption{For the initial state $(0,0,0)$, final population distribution (dark green) according to rotational quantum 
number $N$ defined in \autoref{eq:rot_dist}
after the a) centrifuge and b) Gaussian pulses with peak intensities 
$I_C^0=4.158\cdot10^{10}~$W/cm$^2$ and $I_G^0=10^{12}~$W/cm$^2$, respectively.
For the rigid-rotor approximation (light pink), we present the weights of each rotational state with the lowest vibrational band $\nu=0$.}
\label{fig:nu0_I1e12}
\end{figure}

An experimental optical centrifuge creates molecular
samples in super-rotor states, \ie, in very high rotational excitations~\cite{Korobenko2014}.
We analyze this property by the final cumulative weights in a certain rotational quantum number,
independently of their vibrational band, ${\cal N}_{0,0,0}(N,t=t_f)$ 
defined in~\autoref{eq:rot_dist}. 
 ${\cal N}_{0,0,0}(N,t=t_f)$ is plotted as a function of $N$ 
in~\autoref{fig:nu0_I1e12}~(a) and (b) for the G and C pulses, respectively.
In addition, 
\autoref{fig:nu0_I1e12} shows the final weights
in a certain rotational excitation obtained within the  rigid-rotor description, 
see~\autoref{appendix2}.
The GP-dressed dynamics involves many rotational excitations, 
${\cal N}_{0,0,0}(N,t=t_f)$ shows small-amplitude oscillations 
as $N$ increases and reaches a global maximum at $N=44$, decreasing afterwards.  
The rigid-rotor approximation does not reproduce
the rovibrational results due to the 
large contribution from neighboring vibrational bands. 
For the CP-induced dynamics, 
${\cal N}_{0,0,0}(N,t=t_f)$ also oscillates with $N$ 
with larger amplitudes, and reaches its global 
maximum at $N=36$, approaching zero for higher values of $N$.
For the CP case, rigid-rotor description
provides a good approximation because 
the contributions from higher vibrational bands is rather small
at the weak peak intensity of the CP.
The mean rotational excitations, defined in~\autoref{eq:averageN}, are 
$\langle{\cal N}_{0,0,0}\rangle=32.2$ and $27.6$ for the GP and CP, respectively.
These mean values indicate that the way the energy is transferred by the GP pulse to
the ground state give rise to higher rotational excitations
than in the CP dynamics.

\begin{figure}
\includegraphics[width=0.95\columnwidth]{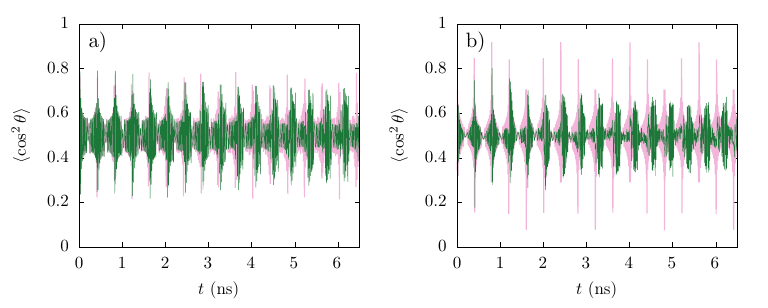}
\caption{For the initial state $(0,0,0)$, time evolution of the alignment (dark-green thin line) 
induced by the a) centrifuge and b) Gaussian pulses with peak intensities $I_G^0=10^{12}~$W/cm$^2$ and 
$I_C^0=4.158\cdot10^{10}~$W/cm$^2$, respectively.
The alignment computed within the rigid-rotor (light-pink thick line) approximation is also plotted.}
\label{fig:nu0_alignment}
\end{figure}

Non-resonant light is normally used to produce samples of 
 molecules aligned along the laser-polarization axis. 
 The time-evolution of the alignment 
$\langle \cos^2 \theta \rangle \equiv \langle \Psi(\bold{R},t) | \cos^2 \theta | \Psi(\bold{R},t)\rangle$
induced by these pulses is presented in~\autoref{fig:nu0_alignment}.
For comparison, the results obtained within the 
rigid-rotor approximation are also plotted in this figure. 
Both pulses provoke an impulsive alignment to the molecule due to their short duration compared to molecular rotational period, 
with $\langle \cos^2 \theta \rangle$ reaching extreme values 
at the rotational revivals $k\tau_B/4$, with $k$ being an integer. 
For the GP-induced alignment, there are large deviations between the
rigid-rotor and rovibrational descriptions, due to the
important role played by higher vibrational bands on the dressed
dynamics.
For instance,
the alignment obtained with the full rovibrational description 
possesses smaller maximal values, which are reduced and shifted with respect to the revivals $k\tau_B/4$ as time increases.
For the CP-induced alignment, the agreement with the rigid-rotor description is good during (approximately) the 
first rotational period. 
Indeed, the narrow vibrational broadening induced by this pulse
provokes that the differences between these results and a small reduction of the maximal alignment
appear only at later times.

\begin{figure}
\includegraphics[width=0.96\columnwidth]{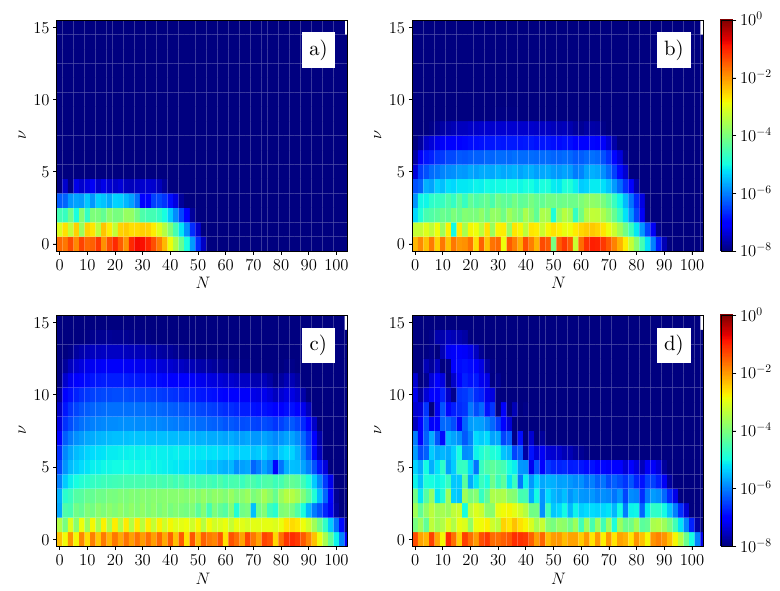}
\caption{For the initial state $(0,0,0)$ in a CP, 
weights $|C_{0,0,0}(\nu,N,t)|^2$ of the field-free rotational and vibrational states 
into the field-dressed wavepacket at times 
a) $t=2.29$~ps (first maximum of $I_C(t)$); 
b) $t=6.86$~ps (fifth maximum);
c) $t=10.97$~ps (twelfth maximum); and
d) at the end of the pulse ($t=15$~ps), with peak intensity  $I_C^0=1.8\cdot10^{11}~$W/cm$^2$.}
\label{fig:2D_0_c_strongI}
\end{figure}

We conclude this section by analyzing the dynamics induced to the initial state $(0,0,0)$ by a CP with 
a stronger intensity, $I_C^0=1.8\cdot10^{11}~$W/cm$^2$, the weights $|C_{0,0,0}(\nu,N,t)|^2$
are plotted in~\autoref{fig:2D_0_c_strongI} at different times.
Due to the stronger peak intensity, already at the first maximum,
a significant population is transferred to the vibrational band $\nu=1$.
By further increasing the time,
the distribution of population becomes significantly wider in $N$, but not in $\nu$.
At the end of the pulse, there are several vibrational bands with significant 
contributions ${\cal V}_{0,0,0}(1,t=t_f)=0.028$ and
${\cal V}_{0,0,0}(2,t=t_f)=0.011$.
However, despite these moderate vibrational weights, 
the CP pulse efficiently transfers population to 
higher rotational excitations
reaching the mean value $\left\langle{\cal N}_{\nu_0,N_0,M_{N_0}} \right\rangle=32.8$.

\subsection{Dynamics of rovibrational excited states}
\label{sec:excited}

\begin{figure}
\includegraphics[scale=0.65]{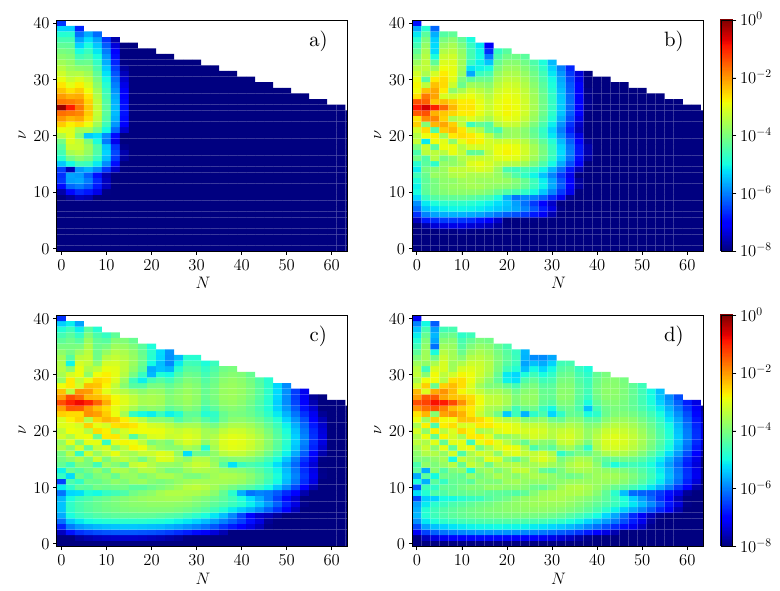}
\caption{For the initial state $(25,0,0)$ in a GP,
weights $|C_{25,0,0}(\nu,N,t)|^2$ of the field-free rotational and vibrational states 
into the field-dressed wavepacket
a) at the FWHM $t=553$~fs; 
b) at the maximum of the GP $t=t_g=672$~fs;
c) at the FWHM $t=790$~fs;
d) and at the end of the GP. The peak intensity is $I_G^0=10^{12}~$W/cm$^2$.}
\label{fig:2D_25_g}
\end{figure}

\begin{figure}
\includegraphics[scale=0.65]{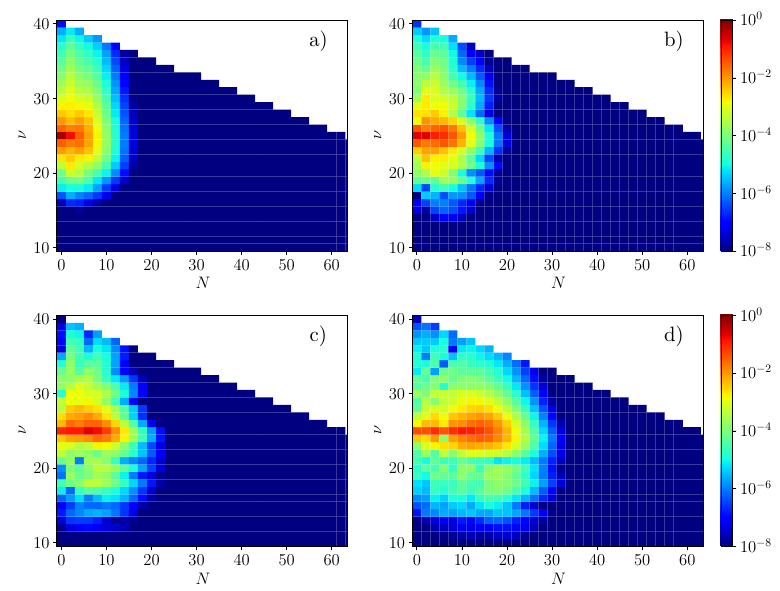}
\caption{For the initial state $(25,0,0)$ in a CP, 
weights $|C_{25,0,0}(\nu,N,t)|^2$ of the field-free rotational and vibrational states into the field-dressed wavepacket at  
a) first maximum $t=2.29$~ps; 
b) fifth maximum $t=6.86$~ps;
c) twelfth maximum $t=10.97$~ps; and 
d) at the end of the pulse $t=15$~ps. Peak intensity is $I_C^0=4.158\cdot10^{10}~$W/cm$^2$.}
\label{fig:2D_25_c}
\end{figure}

In this section, we analyze the impact of these two pulses on excited vibrational states
taking as prototype example the initial state $(25,0,0)$. 
Figs.~\ref{fig:2D_25_g} and~\ref{fig:2D_25_c} present  
the weights $|C_{25,0,0}(\nu,N,t)|^2$ as a function of $\nu$ and $N$ at different
time steps of  the GP- and CP-dressed dynamics, respectively.
The peak intensities are fixed to $I_G^0=10^{12}~$W/cm$^2$ and $I_C^0=4.158\cdot10^{10}~$W/cm$^2$.

At the beginning of the pulses, the population distribution is similar, as illustrated in panels a) of Figs.~\ref{fig:2D_25_g} and~\ref{fig:2D_25_c} for $t=0.553$~ps and $2.29$~ps, respectively. 
Initially, the rotational excitations within $\nu_0=25$ show the largest contribution
to the wavepackets, but those within the neighboring vibrational bands $\nu_0\pm1,2$ also possess significant
weights. 
As time increases, the differences between the field-dressed dynamics become more evident and
are due to the distinct ways that the pulses transfer energy to the molecule. The short duration and high intensity of the GP provoke that significantly more rovibrational states are involved in the dynamics, with a large amount of population transferred to other vibrational bands.
This population distribution to neighboring vibrational bands 
follows a diagonal path due to that the dominant matrix elements between different bands are
$\langle \nu,N,M_N|\Delta\alpha(R)\cos^2\theta|\nu\pm1,N\pm2,M_N\rangle$, see~\autoref{fig:FC}.
For the CP wavepacket, more rotational excitations get populated at these later
time steps, whereas the total weights of states in
different vibrational bands does not change significantly
after the first intensity maximum,
being the main contributions from $\nu_0\pm 1,2$ and $3$.
This controlled spread over vibrational excitations reflects the gradual transfer of energy
from the centrifuge field to the molecule.
Hence, in this strong coupling regime, the CP pulse is more efficient in creating super-rotor wavepackets 
formed by high rotational excitations 
while simultaneously keeping a narrow distribution of vibrational bands.

Compared to the ground state, the impact on this excited state
of these laser fields is
weaker. This is due to the largest spatial extension
of the vibrational wave function as $\nu$ increases, and 
the highest probability density being located close to the outermost classical 
turning point of the electronic potential curve. As a consequence, 
the overlaps of the corresponding wave functions with the $R$-dependent polarizabilities 
are smaller, reducing the impact of the laser field.
In contrast to the ground-state, the CP is more efficient climbing the rotational ladder,
and at the end of the pulses, the mean values of the rotational 
excitations are 
$\left\langle{\cal N}_{25,0,0}\right\rangle=10.35$, and $7.90$, 
for the C and G pulses, respectively. 
For the GP-dressed dynamics, 
the wider distribution of the final weights
in the vibrational and rotational quantum 
numbers  
reduces the contributions of highly excited rotational states,
and, as a consequence,  the mean value
$\left\langle{\cal N}_{25,0,0}\right\rangle$ becomes smaller.

\begin{figure}
\includegraphics[scale=0.65]{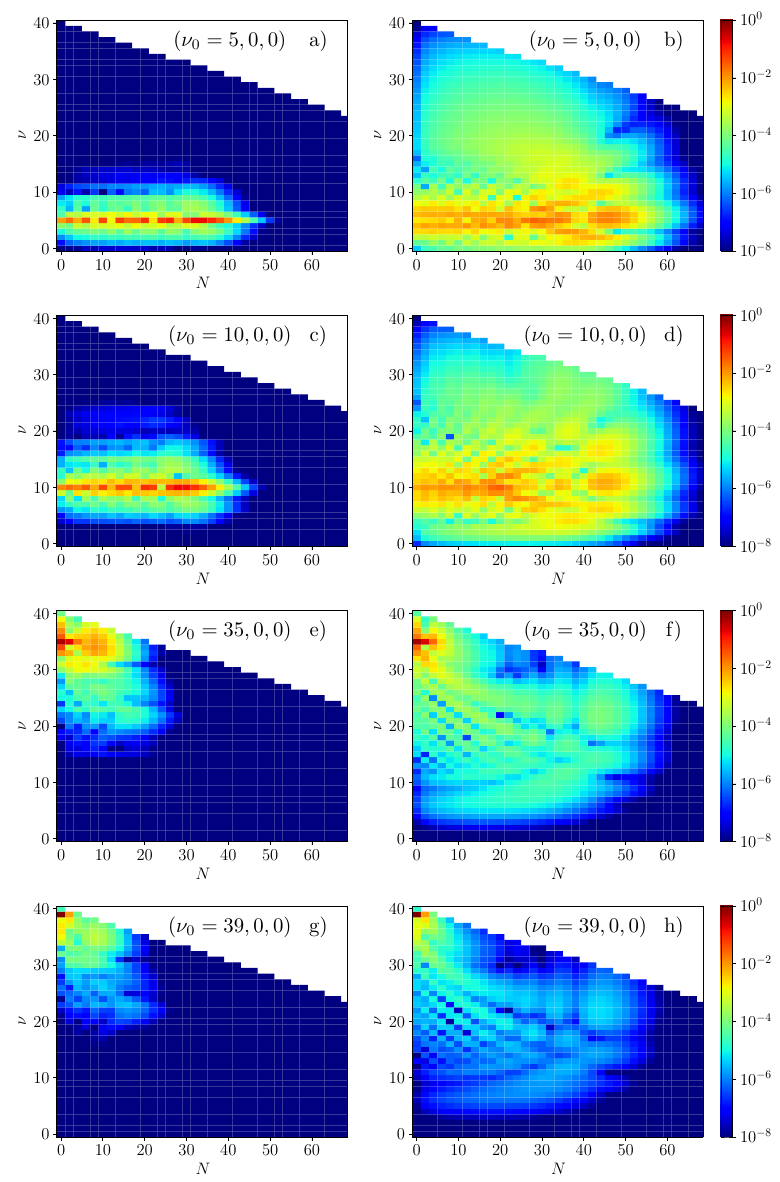}
\caption{For the initial states $(\nu_0,0,0)$, 
weights $|C_{\nu_0,0,0}(\nu,N,t)|^2$ of the field-free rotational and vibrational states 
into the field-dressed wavepacket at the end of the 
centrifuge (left column) and Gaussian (right column)
pulses with  peak intensity $I_G^0=10^{12}~$W/cm$^2$ and
$I_C^0=4.158\cdot10^{10}~$W/cm$^2$, respectively.}
\label{fig:2d_G_nu}
\end{figure}

Analogous results are obtained for
other vibrational excited state 
$(\nu_0,0,0)$, as illustrated in~\autoref{fig:2d_G_nu} with 
the final distribution of population in the rovibrational spectrum 
for $\nu_0=5,\,10,\, 35$ and $39$, and peak intensities
$I_G^0=10^{12}~$W/cm$^2$ and $I_C^0=4.158\cdot10^{10}~$W/cm$^2$.
The dressed dynamics of these states shows similar features as those discussed for $(25,0,0)$ above. 
For both pulses, we observe that as the vibrational excitation increases, the impact of the laser field is reduced.
The stronger impact induced by the GP pulse, gives rise to higher
hybridization of the rotational and vibrational motions illustrated
by wider distributions among the field-free states.
Most important, the
CP-induced vibrational spreading is very narrow for all analyzed cases.

\begin{figure}
\includegraphics[width=0.8\columnwidth]{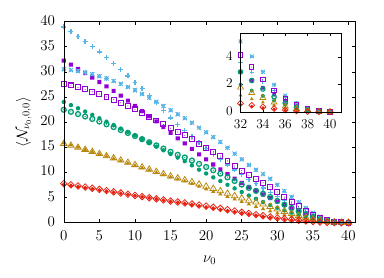}
\caption{The accumulative rotational weights 
at the end of a Gaussian and centrifuge pulse
as a function of the vibrational quantum number of the initial states $(\nu_0,0,0)$. 
The inset shows these accumulative rotational weights 
for initial states close to the dissociation threshold
with  $\nu_0\ge 32$. The peak intensities are fixed so that
both pulses carry the same energy, and are 
$I_C^0=50.00$~GW/cm$^2$ (blue crosses), 
$I_C^0=41.58$~GW/cm$^2$ (purple squares), 
$I_C^0=31.18$~GW/cm$^2$ (green circles), 
$I_C^0=20.79$~GW/cm$^2$ (yellow triangles) and 
$I_C^0=10.39$~GW/cm$^2$ (red diamonds), 
for the centrifuge pulse;
and $I_G^0=1.203$~TW/cm$^2$ (blue stars), 
$I_G^0=1.00$~TW/cm$^2$ (purples filled squares),
$I_G^0=0.75$~TW/cm$^2$ (green filled circles), $I_G^0=0.50$~TW/cm$^2$ (yellow filled triangles) and $I_G^0=0.25$~TW/cm$^2$ (red filled diamonds) for the Gaussian pulse.}
\label{fig:meanN}
\end{figure}

We explore the creation of molecular samples in highly 
excited rotational states by  
the final mean value of the rotational excitation 
$\left\langle{\cal N}_{\nu_0,0,0}\right\rangle$~\autoref{eq:averageN},
which is presented in~\autoref{fig:meanN} as a function of the vibrational excitation 
$\nu_0$ and for several peak intensities of both pulses. 
As indicated above, the impact of the laser field decreases as 
$\nu_0$ increases, which is manifested on the decreasing trend of
$\left\langle{\cal N}_{\nu_0,0,0}\right\rangle$. 
For the lowest peak intensities, $\left\langle{\cal N}_{\nu_0,0,0}\right\rangle$
reaches similar values for both pulses. 
In contrast, for the strongest intensities, 
the rotational ladder  is climbed more (less) efficiently  
by applying a CP than a GP for the rotational ground states in the vibrational bands with 
$\nu_0\gtrsim 14$ ($\nu_0\lesssim 14$).  
This is due to the broader vibrational distribution
induced by the GP for these initial excited states, which
reduces the weight of higher rotational
excitations, and, therefore, the value of 
$\left\langle{\cal N}_{\nu_0,0,0}\right\rangle$. 
For the vibrational bands close to the dissociation threshold, $\left\langle{\cal N}_{\nu_0,0,0}\right\rangle$
reaches rather low values due to the small amount of rotational excitations bounded
in those bands and a weaker field impact. 

\begin{figure}
\includegraphics[width=0.98\columnwidth]{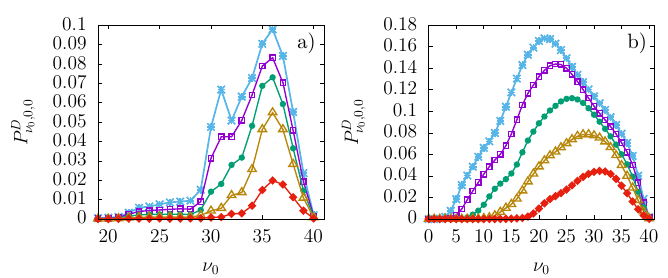}
\caption{For the initial states $(\nu_0,0,0)$, 
population transferred to the 
continuum~\autoref{eq:continuum} at the end of the a) centrifuge and b) Gaussian pulses for several peak intensities: for the centrifuge, $I_C^0=50.00$~GW/cm$^2$ (blue stars), 
$I_C^0=41.58$~GW/cm$^2$ (purple squares), 
$I_C^0=31.18$~GW/cm$^2$ (green filled circles), 
$I_C^0=20.79$~GW/cm$^2$ (yellow triangles) and 
$I_C^0=10.39$~GW/cm$^2$ (red filled diamonds),
and for the Gaussian,
$I_G^0=1.203$~TW/cm$^2$ (blue stars), 
$I_G^0=1.00$~TW/cm$^2$ (purple squares),
$I_G^0=0.75$~TW/cm$^2$ (green filled circles), 
$I_G^0=0.50$~TW/cm$^2$ (yellow triangles) and 
$I_G^0=0.25$~TW/cm$^2$ (red filled diamonds).
These intensities have been taken so that both pulses carry
the same energy, see~\autoref{sec:system}.}
\label{fig:dissociation}
\end{figure}

An important feature of the dressed dynamics is the partial 
dissociation of the molecule as 
scattering states also contribute to the final wave function~\cite{Lemeshko2009}.
This occurs for $(25,0,0)$ in~\autoref{fig:2D_25_c}~(a),
and for $\nu_0=5,10,35$ and $39$ under
the CP in~\autoref{fig:2d_G_nu}. 
For all these states, 
we encounter that the upper most rotational
excitation bounded within a certain vibrational band is populated. 
This field-induced dissociation at the end of the pulse is quantified $P^D_{\nu_0,0,0}$, defined in~\autoref{eq:continuum},
which is plotted in~\autoref{fig:dissociation}~(a) and~(b)
as a function of $\nu_0$ and for several peak intensities of
the G and C pulses, respectively. 
This field-induced dissociation strongly depends on the initial 
state, pulse shape and peak intensity.
For the GP, $P^D_{\nu_0,0,0}$ shows a smooth behavior as a function of 
the vibrational band of the initial state $\nu_0$.
$P^D_{\nu_0,0,0}$ initially increases as $\nu_0$ increases, 
reaching a maximum, and decreasing for highly excited vibrational bands. 
As the peak intensity increases, a largest amount of population is transferred to the continuum,
and, in addition, the maximum is shifted to lower values of $\nu_0$, being
at $\nu_0=36$ and $\nu_0=22$ for $I_G^0=2.5\cdot 10^{11}$~W/cm$^2$ and $1.203\cdot 10^{12}$~W/cm$^2$,
respectively.
For the CP, the dependence of $P^D_{\nu_0,0,0}$ on $\nu_0$ changes as $I_C^0$ increases, transforms from having only one maximum to two, being the global one located at $\nu_0=36$ for all considered peak intensities $I_G^0$. 

The narrower vibrational spreading induced by the CP pulse favors the reduction of 
the dissociation for most initial states. 
Indeed, the stronger vibrational impact of the GP is reflected in a higher value of $P^D_{\nu_0,0,0}$ for most of the
analyzed states, except for the highly excited vibrational ones lying
close to the dissociation limit. 
For instance, for $\nu_0=36$, $P^D_{\nu_0,0,0}=0.06$ and $0.08$
for a GP with $I_G^0=10^{12}~$W/cm$^2$ and 
a CP with $I_C^0=4.158\cdot10^{10}~$W/cm$^2$, respectively.
For the GP, the ladder-like distribution of population implies that lower-lying 
vibrational bands are populated, and these bands accommodate more rotational excitations,
being, therefore, harder the dissociation. 
In contrast, at the end of CP, mainly 
neighboring vibrational bands to $\nu_0=36$ are populated, being easier
to reach their maximal rotational excitation and
facilitating the molecular dissociation.

\subsection{Thermal distribution}

The results discussed in the previous sections assume either a 
rotational temperature of $0$~K, \ie, the $(0,0,0)$ state,  
or vibrational excitations but no rotational excitation $N_0=0$.
Molecular samples in a single rovibrational state are experimentally feasible in the ultracold regime~\cite{Kangkuen2010,simon2014}. 
However, molecular beam experiments are characterized
by a certain rotational temperature, 
being extremely challenging to create
molecular samples in a single state~\cite{jochen2009,jochhen2009_2,Kienitz2016}.
Here, we consider thermal samples of Rb$_2$ with rotational
temperatures that are experimentally feasible $T\leq2.0~$K.  
In our description, the thermal sample is restricted to
states within the lowest vibrational band, \ie, with quantum numbers 
$(0,N,M_N)$ with $N \leq 24$ and $N\leq M_N\leq N$.
Note that for $T=2~$K, the weight of field-free states 
with rotational quantum number $N>24$ 
is smaller than $0.01$. 

\begin{figure}
\includegraphics[width=0.98\columnwidth]{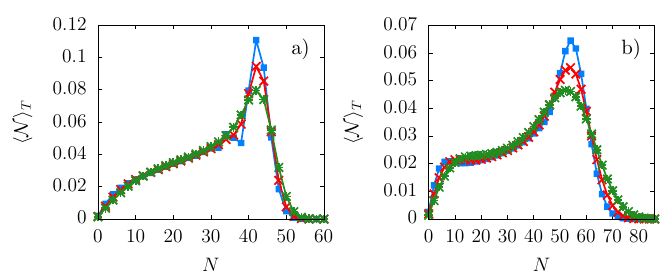}
\caption{
For thermal samples of Rb$_2$ at rotational
temperatures $T=0.5$~K (blue squares), $T=1.0$~K (red crosses) and $T=2.0$~K (green stars), 
post-pulse distribution of the population among the rotational states~\autoref{eq:averageN} for
a) a Centrifuge pulse with $I_C^0=5.0\cdot10^{10}~$W/cm$^2$, and b) a Gaussian pulse with peak intensity $I_G^0=1.203\cdot10^{12}~$W/cm$^2$. }
\label{fig:thermal}
\end{figure}

In~\autoref{fig:thermal}, we present the thermal weights of
the rotational excitations $\langle{\cal N}\rangle_T$, defined 
in~\autoref{eq:averageN-thermal}, at the end of the Gaussian 
and centrifuge pulses with peak intensities 
$I_G^0=1.203\cdot10^{12}~$W/cm$^2$ and 
$I_C^0=5.0\cdot10^{10}~$W/cm$^2$, respectively.
Note that at these intensities and for
the initial states $(0,N,M_N)$, no population is transferred
to the continuum. 
For both pulses, these thermal distributions 
are very similar. They show an increasing trend till 
reach a maximum for $N=54$ and $42$ for the GP and CP, respectively.
This demonstrates the creation of samples in highly excited
rotational states such as super-rotors.
As the temperature increases, these maxima are reduced 
due to  the larger weights of higher rotational excitations 
to the thermal sample, for which the impact of the external fields is smaller.

\section{Conclusions}
\label{sec:conclu}
We have explored the ro-vibrational dynamics of Rb$_2$ molecules in the $a^3\Sigma^+$ electronic state, due to non-resonant light. 
For the laser field, we have 
considered two different time envelopes having the same spectral bandwidth and carrying the same amount of energy, but
inducing different dynamics. 
For experimentally feasible laser intensities, 
we find a strong coupling between the vibrational and rotational degrees of freedom. Despite the breakdown of the rigid-rotor approximation, these laser fields create wavepackets characterized by very high angular momenta. 
We find that the creation of these highly excited rotational excitations can result in the dissociation of 
the molecule even when initially in deeply bound states, in line with an earlier prediction~\cite{Spanner2001}.

For both pulses, the lack of selection rules in the vibrational quantum number hinders the control over 
the vibrational bands being populated. As a consequence, the field-dressed dynamics involve states from 
lower and upper lying vibrational bands, with similar weights. 
However, a comparison of the dynamics for both pulses shows that the centrifuge one enables a better
control over the vibrational excitations creating narrower distribution in this degree of freedom.
Thus, the centrifuge field provokes the excitation of 
high rotational states while maintaining relatively low spread along the vibrational bands.
This partial control over the vibrational excitations is due to 
the gradual transfer of energy from the centrifuge pulse to the molecule.
Initially, when the centrifuge intensity is increased till the first maximum, 
a few neighboring vibrational bands get weakly populated, but 
this vibrational-rotational coupling is not significantly enhanced in the consecutive maxima.
As a consequence, the centrifuge field is more efficient in creating super-rotor wavepackets with low
impact on the vibrational motion. 
In contrast,  the shorter duration of the Gaussian pulse implies that the energy is transferred faster to the molecule, and results in a stronger coupling between the vibrational and rotational motions. 
Thus, the amount of population that is accumulated in higher rotational excitations might be reduced, due to the larger role played by neighboring vibrational bands in the field-dressed dynamics.

The results of our work offer new perspectives on the utility of the optical centrifuge to control not only rotational, but also vibrational molecular dynamics. Our results provide guidance in, and we hope will advance, those ultrafast spectroscopic studies in which separating the rotational and vibrational excitations by intense laser pulses is of importance.

\appendix

\section{Rigid-Rotor approximation}
\label{appendix2}
Within the rigid-rotor approximation, 
Hamiltonian of a diatomic molecule in a
non-resonant laser pulse linearly polarized along the LFF $Z$-axis,
is given by 
\begin{equation}
{H^{R}}(t)=
\frac{\bold{N}^2}{2\mu }
\langle R^{-2}\rangle_\nu+H_I^{R}(t)
\label{eq:Htotal_rr}
\end{equation}
where the first stands for the rotational kinetic energies, 
with the rotational constant given by 
$B_\nu=\frac{\hbar^2 \langle R^{-2}\rangle_\nu}{2\mu}$.
The interaction of the non-resonant light with
the  polarizability reads
\begin{equation}
H_I^{R}(t)=
-\frac{I(t)}{2c\varepsilon_0}\left(\langle\Delta\alpha(R)\rangle_\nu\cos^2\theta+
\langle\alpha_\perp(R)\rangle_\nu\right).
\label{eq:Hinter_rr}
\end{equation}
The matrix elements 
$\langle f(R)\rangle_\nu= \langle \phi_{\nu,0}|f(R)|\phi_{\nu,0}\rangle_\nu$,
with $f(R)=R^{-2},\, \Delta\alpha(R)$ and $\alpha_\perp(R)$,
and $\phi_{\nu,0}=\phi_{\nu,0}(R)$ the field-free vibrational 
wave function, adapt this approximation to the different 
vibrational bands of the 
molecular spectrum~\cite{gonzalez2004}. 
The time-dependent Schrödinger equation associated  with Hamiltonian~\autoref{eq:Htotal_rr} is solved by 
the short iterative Lanczos algorithm for the time propagation~\cite{beck20001} and a 
basis set expansion in terms of the spherical Harmonics for the angular coordinates~\cite{omiste2013}.

\section{Weights and matrix elements}
\label{appendix1}

The field-dressed dynamics is analyzed by projecting the wave packet onto the basis formed
by the field-free eigenstates of the electronic state $a^3\Sigma^+$ of Rb$_2$ as in~\autoref{eq:expan}.
For a given initial state $(\nu_0, N_0, M_{N_0})$,
the weight of the eigenstate $(\nu, N, M_{N_0})$ to the field-dressed wave function is 
$|C_{\nu_0, N_0,M_{N_0}} (\nu, N,t)|^2$. 
To illustrate the breakdown of the rigid-rotor approximation, we 
calculate the population transferred to a certain vibrational band 
$\nu$ as
\begin{equation}
{\cal V}_{\nu_0,N_0,M_{N_0}}(\nu,t)=
\sum_{N=0}^{N_{max}}|C_{\nu_0, N_0,M_{N_0}} (\nu, N,t))|^2
\label{eq:vib_dist}
\end{equation}
where $N_{max}$ is the maximum number of rotational states included 
in the basis set expansion~\eqref{eq:expan}. 
Analogously, an accumulative rotational distribution, independent of the vibrational excitations, can be defined as 
\begin{equation}
{\cal N}_{\nu_0,N_0,M_{N_0}}(N,t)=\sum_{\nu}
|C_{\nu_0, N_0,M_{N_0}} (\nu, N,t)|^2\, 
\label{eq:rot_dist}
\end{equation}
where the sum runs over all vibrational bands of the Rb$_2$ $a^3\Sigma^+$ electronic state.
At the of the pulses, we compute the mean value of the rotational excitations defined as 
\begin{equation}
\left\langle
{\cal N}_{\nu_0,N_0,M_{N_0}}
\right\rangle=\frac{\sum_{N=0}^{N_{max}} N {\cal N}_{\nu_0,N_0,M_{N_0}}(N,t_f)}
{\sum_{N=0}^{N_{max}} {\cal N}_{\nu_0,N_0,M_{N_0}}(N,t_f)}.
\label{eq:averageN}
\end{equation}  
with $t_f=t_G$ and $t_C$ for the GP and CP, respectively. 
These expressions~\ref{eq:vib_dist},~\ref{eq:rot_dist} and~\ref{eq:averageN} take into
account that $M_{N_0}$ is a good quantum number due to the azimuthal symmetry.

Due to the strong field impact, part of the population is transferred to scattering states,
\ie, to the continuum. This effect is quantified by 
\begin{equation}
P^D_{\nu_0, N_0,M_{N_0}}=1-
\sum_{\nu,N}
|C_{\nu_0, N_0,M_{N_0}} (\nu, N,t_f)|^2\, 
\label{eq:continuum}
\end{equation}
at the end of the G and C pulses, \ie,  $t_f=t_G$ and $t_C$, respectively. 
The sum in~\autoref{eq:continuum} rums over all vibrational bands, and all rotational excitations
included in the numerical treatment. 

For a molecular sample at temperature $T$, the 
thermal average of the rotational excitations reads as 
\begin{equation}
\left\langle 
{\cal N}\right\rangle_T=
\sum_{\nu_0,N_0}
P_{\nu_0,N_0,M_{N_0}}(N,t_f) W_{\nu_0N_0|M_{N_0}|},
\label{eq:averageN-thermal}
\end{equation}
with $t_f=t_G$ and $t_C$ for the GP and CP, respectively. 
The Maxwell-Boltzmann weights are
\begin{equation}
W_{\nu_0N_0|M_{N_0}|}=\frac{g_{|M_{N_0}|}e^\frac{E_{0,0}-E_{\nu_0,N_0}}{Tk_B}}{Z},
\label{eq:weights_thermal}
\end{equation}
where $k_B$ is the Boltzmann constant, $E_{0,0}$  and $E_{\nu_0,N_0}$ are the field-free energies
of the $(0,0,0)$  and $(\nu_0,N_0,M_{N_0})$ states, respectively.
The factor $g_{M_{N_0}}$  
takes into account that the impact of the laser field on the states
$(\nu_0,N_0,|M_{N_0}|)$ and $(\nu_0,N_0,-|M_{N_0}|)$ is identical, so that 
$g_{0}=1$ and 
$g_{|M_{N_0}|}=2$ for  $|M_{N_0}|\ne 0$.
The normalization constant is given by 
\begin{equation}
Z=
\sum_{N_0=0}^{N_T} (2N_0+1)e^\frac{E_{0,0}-E_{\nu_0,N_0}}{Tk_B},
\label{eq:weights_thermal_Z}
\end{equation}
where $(2N_0+1)$
is the field-free degeneracy in the magnetic quantum number.
The highest 
rotational excitation included  is $N_T\le 24$.
For the considered temperatures, we neglect the contribution 
of states in excited vibrational bands, restricting 
$\nu_0=0$ in~\autoref{eq:weights_thermal} and~\autoref{eq:weights_thermal_Z}.
Indeed, the thermal weight for the state $(1,0,0)$, \ie, the rotational ground state of the
first excited vibrational band, is 
$W_{1,0,0}=5.9\cdot10^{-18}$ and $1.5\cdot10^{-6}$, for $T=0.5$~K and $T=2$~K, respectively.

\begin{figure}[H]
    \centering
    \includegraphics[width=0.98\linewidth]{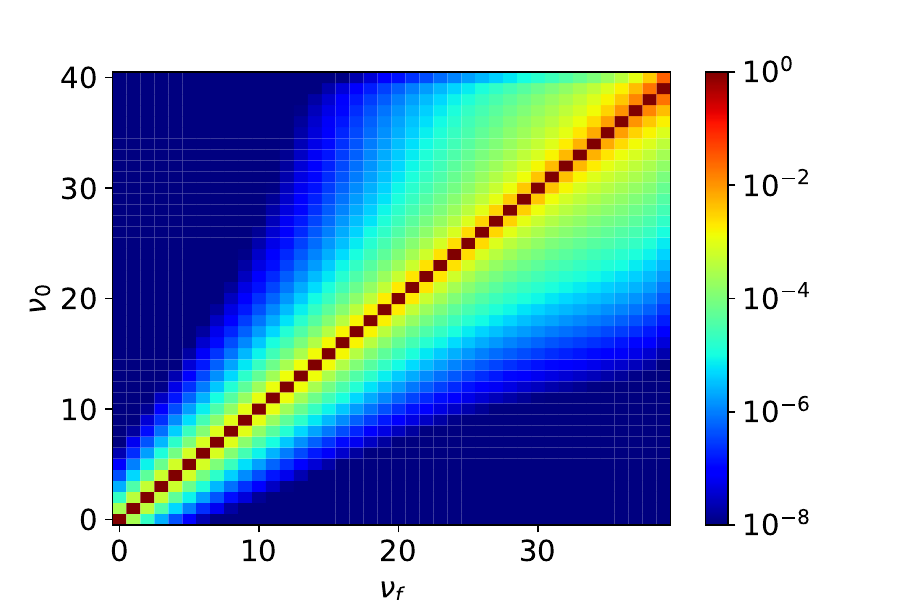}
    \caption{Matrix elements of the vibrational part of the interaction~\autoref{eq:alpha_me} between the 
    states $(\nu_0,0,0)$ and $(\nu_f,2,0)$.}
    \label{fig:FC}
\end{figure}

To better understand the mechanism behind the population transferred to other vibrational 
bands,~\autoref{fig:FC} shows the radial matrix elements
\begin{equation}
\label{eq:alpha_me}
\langle \nu,0
|\Delta\alpha(R)|\nu', 2
\rangle =\int\phi^*_{\nu,0}(R)\Delta\alpha(R)\phi_{\nu',2}(R) R^2dR,
\end{equation}
for the non-diagonal terms in the angular operator $\cos^2\theta$,
\ie, $\Delta N=\pm 2$. 

\section*{Acknowledgements}

Financial support by the Spanish projects PID2023-147039NB-I00 (MICIN), the Andalusian research group FQM-207 and the Deutsche
Forschungsgemeinschaft through the joint ANR-DFG CoRoMo Projects No. 505622963 (KO 2301/15-1). 
J.M.G.G. also acknowledges financial support by the grant 
PRE2021-099603 funded by MICIU/AEI/10.13039/501100011033 and by “ESF+”.

\bibliography{biblio} 

@article{Chen2023,
	author = {Chen, Timothy Y. and Steinmetz, Scott A. and Patterson, Brian D. and Jasper, Ahren W. and Kliewer, Christopher J.},
	date = {2023/06/03},
	doi = {10.1038/s41467-023-38873-z},
	id = {Chen2023},
	isbn = {2041-1723},
	journal = {Nat. Commun.},
	number = {1},
	pages = {3227},
	title = {Direct observation of coherence transfer and rotational-to-vibrational energy exchange in optically centrifuged CO2 super-rotors},
	url = {https://doi.org/10.1038/s41467-023-38873-z},
	volume = {14},
	year = {2023}}

@Article{jochhen2009_2,
author ={Nevo, Iftach and Holmegaard, Lotte and Nielsen, Jens H. and Hansen, Jonas L. and Stapelfeldt, Henrik and Filsinger, Frank and Meijer, Gerard and K\"upper, Jochen},
title  ="Laser-induced 3D alignment and orientation of quantum state-selected molecules",
journal  ="Phys. Chem. Chem. Phys.",
year  ="2009",
volume  ="11",
issue  ="42",
pages  ="9912-9918",
publisher  ="The Royal Society of Chemistry",
doi  ="10.1039/B910423B",
url  ="http://dx.doi.org/10.1039/B910423B"}

@article{jochen2009,
  title = {Laser-Induced Alignment and Orientation of Quantum-State-Selected Large Molecules},
  author = {Holmegaard, Lotte and Nielsen, Jens H. and Nevo, Iftach and Stapelfeldt, Henrik and Filsinger, Frank and K\"upper, Jochen and Meijer, Gerard},
  journal = {Phys. Rev. Lett.},
  volume = {102},
  issue = {2},
  pages = {023001},
  numpages = {4},
  year = {2009},
  month = {Jan},
  publisher = {American Physical Society},
  doi = {10.1103/PhysRevLett.102.023001},
  url = {https://link.aps.org/doi/10.1103/PhysRevLett.102.023001}
}

@article{Spanner2001, 
    author = {Spanner, M. and Ivanov, M. Yu.},
    title = {Angular trapping and rotational dissociation of a diatomic molecule in an optical centrifuge},
    journal = {J. Chem. Phys.},
    volume = {114},
    number = {8},
    pages = {3456-3464},
    year = {2001},
    month = {02},
    issn = {0021-9606},
    doi = {10.1063/1.1342225},
    url = {https://doi.org/10.1063/1.1342225},
}

@article{Kienitz2016,
author = {Kienitz, Jens S. and Trippel, Sebastian and Mullins, Terry and Dlugolecki, Karol and González-Férez, Rosario and K\"upper, Jochen},
title = {Adiabatic Mixed-Field Orientation of Ground-State-Selected Carbonyl Sulfide Molecules},
journal = {ChemPhysChem},
volume = {17},
number = {22},
year = {2016},
pages = {3740-3746},
doi = {https://doi.org/10.1002/cphc.201600710}
}

@article{Kangkuen2010,
  title = {Controlling the Hyperfine State of Rovibronic Ground-State Polar Molecules},
  author = {Ospelkaus, S. and Ni, K.-K. and Qu\'em\'ener, G. and Neyenhuis, B. and Wang, D. and de Miranda, M. H. G. and Bohn, J. L. and Ye, J. and Jin, D. S.},
  journal = {Phys. Rev. Lett.},
  volume = {104},
  issue = {3},
  pages = {030402},
  numpages = {4},
  year = {2010},
  month = {Jan},
  publisher = {American Physical Society},
  doi = {10.1103/PhysRevLett.104.030402},
  url = {https://link.aps.org/doi/10.1103/PhysRevLett.104.030402}
}

@article{simon2014,
  title = {Creation of Ultracold $^{87}\mathrm{Rb}^{133}\mathrm{Cs}$ Molecules in the Rovibrational Ground State},
  author = {Molony, Peter K. and Gregory, Philip D. and Ji, Zhonghua and Lu, Bo and K\"oppinger, Michael P. and Le Sueur, C. Ruth and Blackley, Caroline L. and Hutson, Jeremy M. and Cornish, Simon L.},
  journal = {Phys. Rev. Lett.},
  volume = {113},
  issue = {25},
  pages = {255301},
  numpages = {5},
  year = {2014},
  month = {Dec},
  publisher = {American Physical Society},
  doi = {10.1103/PhysRevLett.113.255301},
  url = {https://link.aps.org/doi/10.1103/PhysRevLett.113.255301}
}

@article{gonzalez2004,
  title = {Rovibrational spectra of diatomic molecules in strong electric fields: The adiabatic regime},
  author = {Gonz\'alez-F\'erez, R. and Schmelcher, P.},
  journal = {Phys. Rev. A},
  volume = {69},
  issue = {2},
  pages = {023402},
  numpages = {11},
  year = {2004},
  month = {Feb},
  publisher = {American Physical Society},
  doi = {10.1103/PhysRevA.69.023402},
  url = {https://link.aps.org/doi/10.1103/PhysRevA.69.023402}
}

@article{Gonzalezferez2012,
  title = {Enhancing photoassociation rates by nonresonant-light control of shape resonances},
  author = {Gonz\'alez-F\'erez, Rosario and Koch, Christiane P.},
  journal = {Phys. Rev. A},
  volume = {86},
  issue = {6},
  pages = {063420},
  numpages = {6},
  year = {2012},
  month = {Dec},
  publisher = {American Physical Society},
  doi = {10.1103/PhysRevA.86.063420},
  url = {https://link.aps.org/doi/10.1103/PhysRevA.86.063420}
}

@article{salas2023,
  title = {Nonlinear dynamics of molecular superrotors},
  author = {Chandre, C. and Salas, J. Pablo},
  journal = {Phys. Rev. A},
  volume = {107},
  issue = {6},
  pages = {063105},
  numpages = {12},
  year = {2023},
  month = {Jun},
  publisher = {American Physical Society},
  doi = {10.1103/PhysRevA.107.063105},
  url = {https://link.aps.org/doi/10.1103/PhysRevA.107.063105}
}

@article{Lemeshko2009,
  title = {Probing Weakly Bound Molecules with Nonresonant Light},
  author = {Lemeshko, Mikhail and Friedrich, Bretislav},
  journal = {Phys. Rev. Lett.},
  volume = {103},
  issue = {5},
  pages = {053003},
  numpages = {4},
  year = {2009},
  month = {Jul},
  publisher = {American Physical Society},
  doi = {10.1103/PhysRevLett.103.053003},
  url = {https://link.aps.org/doi/10.1103/PhysRevLett.103.053003}
}

@article{Karczmarek1999,
  title = {Optical Centrifuge for Molecules},
  author = {Karczmarek, Joanna and Wright, James and Corkum, Paul and Ivanov, Misha},
  journal = {Phys. Rev. Lett.},
  volume = {82},
  issue = {17},
  pages = {3420--3423},
  numpages = {0},
  year = {1999},
  month = {Apr},
  publisher = {American Physical Society},
  doi = {10.1103/PhysRevLett.82.3420},
  url = {https://link.aps.org/doi/10.1103/PhysRevLett.82.3420}
}

@article{Korobenko2014,
  title = {Direct Observation, Study, and Control of Molecular Superrotors},
  author = {Korobenko, Aleksey and Milner, Alexander A. and Milner, Valery},
  journal = {Phys. Rev. Lett.},
  volume = {112},
  issue = {11},
  pages = {113004},
  numpages = {5},
  year = {2014},
  month = {Mar},
  publisher = {American Physical Society},
  doi = {10.1103/PhysRevLett.112.113004},
  url = {https://link.aps.org/doi/10.1103/PhysRevLett.112.113004}
}

@article{beck20001,
title = {The multiconfiguration time-dependent Hartree (MCTDH) method: a highly efficient algorithm for propagating wavepackets},
journal = {Phys. Rep.},
volume = {324},
number = {1},
pages = {1-105},
year = {2000},
issn = {0370-1573},
doi = {https://doi.org/10.1016/S0370-1573(99)00047-2},
url = {https://www.sciencedirect.com/science/article/pii/S0370157399000472},
author = {M.H. Beck and A. Jäckle and G.A. Worth and H.-D. Meyer}
}

@article{omiste2013,
  title = {Rotational dynamics of an asymmetric-top molecule in parallel electric and nonresonant laser fields},
  author = {Omiste, Juan J. and Gonz\'alez-F\'erez, Rosario},
  journal = {Phys. Rev. A},
  volume = {88},
  issue = {3},
  pages = {033416},
  numpages = {11},
  year = {2013},
  month = {Sep},
  publisher = {American Physical Society},
  doi = {10.1103/PhysRevA.88.033416},
  url = {https://link.aps.org/doi/10.1103/PhysRevA.88.033416}
}

@Article{RonnieReview88,
  author = 	 {Kosloff, R.},
  title = 	 "{Time-Dependent Quantum-Mechanical Methods for Molecular Dynamics}",
  journal = 	 {J. Phys. Chem.},
  year = 	 {1988},
  volume = 	 {92},
  pages = 	 {2087-2100},
url = {https://doi.org/10.1021/j100319a003},
doi = {10.1021/j100319a003},
  optnote = 	 {Representation of wave function and operators, i.e. grid rep. and Fourier/pseudospectral method, Propagation methods: Chebychev, Lanczos, second order differentiating, split op. (incl. spectrum of
                  corr. fct. from Chebychev, time-dep. Hamiltonians)}
}

@Article{RonnieReview94,
  author = 	 {Kosloff, R.},
  title = 	 {Propagation methods for molecular dynamics},
  journal = 	 {Annu. Rev. Phys. Chem.},
  year = 	 {1994},
  volume = 	 {45},
  pages = 	 {145-178},
url ={https://doi.org/10.1146/annurev.pc.45.100194.001045},
doi ={10.1146/annurev.pc.45.100194.001045},
  optnote =         {Review of Newton and Chebychevpropagator for Hermitian and Nonhermitian problems,Chebychev for Schr\"odinger eq., eigenfunctions,
		  Green's fct., Correlation fcts.}
}

@article{LightCarrington,
author = {Light, John C. and Carrington, Tucker},
title = {Discrete-Variable Representations and their Utilization},
journal = {Adv. Chem. Phys.},
volume = 114,
year = {2007},
pages = {263--310},
publisher = {John Wiley & Sons, Inc.},
isbn = {9780470141731},
url = {http://dx.doi.org/10.1002/9780470141731.ch4},
doi = {10.1002/9780470141731.ch4}
}

@Article{KochRMP19,
  title = {Quantum control of molecular rotation},
  author = {Koch, Christiane P. and Lemeshko, Mikhail and Sugny, Dominique},
  journal = {Rev. Mod. Phys.},
  volume = {91},
  pages = {035005},
  year = {2019},
  month = {Sep},
    OPTnote = 	 {arXiv:1810.11338},
  doi = {10.1103/RevModPhys.91.035005},
  url = {https://link.aps.org/doi/10.1103/RevModPhys.91.035005}
}

@article{Stapelfeldt2003,
   Author = {Stapelfeldt, H. and Seideman, T.},
   Title = {\textit{Colloquium} : Aligning molecules with strong laser pulses},
   Journal = {Rev. Mod. Phys.},
   Volume = {75},
   Number = {2},
   Pages = {543-557},
      Year = {2003} }

@article{Ohshima2010,
   Author = {Ohshima, Y. and Hasegawa, H.},
   Title = {Coherent rotational excitation by intense nonresonant laser fields},
   Journal = {Int. Rev. Phys. Chem.},
   Volume = {29},
   Number = {4},
   Pages = {619-663},
      Year = {2010} }

@article{Fleischer2012,
   Author = {Fleischer, S. and Khodorkovsky, Y. and Gershnabel, E. and Prior, Y. and Averbukh, I. S.},
   Title = {Molecular Alignment Induced by Ultrashort Laser Pulses and Its Impact on Molecular Motion},
   Journal = {Isr. J. Chem.},
   Volume = {52},
   Number = {5},
   pages = {414-437},
      Year = {2012} }

@article{
Villeneuve2000,
   Author = {Villeneuve, D. M. and Aseyev, S. A. and Dietrich, P. and Spanner, M. and Ivanov, M. Y. and Corkum, P. B.},
   Title = {Forced Molecular Rotation in an Optical Centrifuge},
   Journal = {Phys. Rev. Lett.},
   Volume = {85},
   Number = {3},
   Pages = {542--545},
      Year = {2000} }

@article{
MacPhail2020,
   Author = {MacPhail-Bartley, I. and Wasserman, W. W. and Milner, A. A. and Milner, V.},
   Title = {Laser control of molecular rotation: Expanding the utility of an optical centrifuge},
   Journal = {Rev. Sci. Instrum.},
   Volume = {91},
   Number = {4},
   Pages = {045122},
      Year = {2020} }

@article{
Mullin2025,
   Author = {Mullin, A. S.},
   Title = {Generating Superrotors and Dynamics of Molecules in Extremely High Rotational States},
   Journal = {Annu. Rev. Phys. Chem.},
   Volume = {76},
   Number = {Volume 76, 2025},
   Pages = {357-377},
url = {https://doi.org/10.1146/annurev-physchem-082423-012311},
doi = {10.1146/annurev-physchem-082423-012311},
   Year = {2025} }

@article{LemehskoPRL2009,
  title = {Probing Weakly Bound Molecules with Nonresonant Light},
  author = {Lemeshko, Mikhail and Friedrich, Bretislav},
  journal = {Phys. Rev. Lett.},
  volume = {103},
  issue = {5},
  pages = {053003},
  numpages = {4},
  year = {2009},
  month = {Jul},
  publisher = {American Physical Society},
  doi = {10.1103/PhysRevLett.103.053003},
  url = {https://link.aps.org/doi/10.1103/PhysRevLett.103.053003}
}

@article{LangPRL2008,
  title = {Ultracold Triplet Molecules in the Rovibrational Ground State},
  author = {Lang, F. and Winkler, K. and Strauss, C. and Grimm, R. and Denschlag, J. Hecker},
  journal = {Phys. Rev. Lett.},
  volume = {101},
  issue = {13},
  pages = {133005},
  numpages = {4},
  year = {2008},
  month = {Sep},
  publisher = {American Physical Society},
  doi = {10.1103/PhysRevLett.101.133005},
  url = {https://link.aps.org/doi/10.1103/PhysRevLett.101.133005}
}

@article{DeissPRL2014,
  title = {Probing the Axis Alignment of an Ultracold Spin-polarized ${\mathrm{Rb}}_{2}$ Molecule},
  author = {Dei\ss{}, Markus and Drews, Bj\"orn and Deissler, Benjamin and Hecker Denschlag, Johannes},
  journal = {Phys. Rev. Lett.},
  volume = {113},
  issue = {23},
  pages = {233004},
  numpages = {5},
  year = {2014},
  month = {Dec},
  publisher = {American Physical Society},
  doi = {10.1103/PhysRevLett.113.233004},
  url = {https://link.aps.org/doi/10.1103/PhysRevLett.113.233004}
}

@article{WolfPRL2019,
  title = {Hyperfine Magnetic Substate Resolved State-to-State Chemistry},
  author = {Wolf, Joschka and Dei\ss{}, Markus and Hecker Denschlag, Johannes},
  journal = {Phys. Rev. Lett.},
  volume = {123},
  issue = {25},
  pages = {253401},
  numpages = {5},
  year = {2019},
  month = {Dec},
  publisher = {American Physical Society},
  doi = {10.1103/PhysRevLett.123.253401},
  url = {https://link.aps.org/doi/10.1103/PhysRevLett.123.253401}
}

@article{HeScience2020,
author = {Xiaodong He  and Kunpeng Wang  and Jun Zhuang  and Peng Xu  and Xiang Gao  and Ruijun Guo  and Cheng Sheng  and Min Liu  and Jin Wang  and Jiaming Li  and G. V. Shlyapnikov  and Mingsheng Zhan },
title = {Coherently forming a single molecule in an optical trap},
journal = {Science},
volume = {370},
number = {6514},
pages = {331-335},
year = {2020},
doi = {10.1126/science.aba7468},
URL = {https://www.science.org/doi/abs/10.1126/science.aba7468},
eprint = {https://www.science.org/doi/pdf/10.1126/science.aba7468}
}

@article{Li2000,
   author = {Li, Jing and Bahns, John T. and Stwalley, William C.},
   title = {Scheme for state-selective formation of highly rotationally excited diatomic molecules},
   journal = {J. Chem. Phys.},
   volume = {112},
   number = {14},
   pages = {6255-6261},
   url = {http://dx.doi.org/10.1063/1.481189},
   year = {2000},
   type = {Journal Article}
}

@article{Milner2016,
  title = {Field-free long-lived alignment of molecules with a two-dimensional optical centrifuge},
  author = {Milner, A. A. and Korobenko, A. and Milner, V.},
  journal = {Phys. Rev. A},
  volume = {93},
  issue = {5},
  pages = {053408},
  numpages = {7},
  year = {2016},
  month = {May},
  publisher = {American Physical Society},
  doi = {10.1103/PhysRevA.93.053408},
  url = {https://link.aps.org/doi/10.1103/PhysRevA.93.053408}
}
\bibliographystyle{rsc} 
\end{document}